\documentclass{SciPost}

\hypersetup{
    colorlinks,
    linkcolor={red!50!black},
    citecolor={blue!50!black},
    urlcolor={blue!80!black}
}

\usepackage{graphicx}%
\usepackage{multirow}%
\usepackage{amsmath}%
\usepackage{xcolor}%
\usepackage{textcomp}%
\usepackage{manyfoot}%
\usepackage{booktabs}%
\DeclareMathOperator{\tr}{tr}
\DeclareMathOperator{\ad}{ad}
\DeclareMathOperator{\Ad}{Ad}
\DeclareMathOperator{\arcosh}{arcosh}
\usepackage[capitalise]{cleveref}
\usepackage{booktabs}

\usepackage[bitstream-charter]{mathdesign}
\DeclareSymbolFont{usualmathcal}{OMS}{cmsy}{m}{n}
\DeclareSymbolFontAlphabet{\mathcal}{usualmathcal}

\fancypagestyle{SPstyle}{
\fancyhf{}
\lhead{\colorbox{scipostblue}{\bf \color{white} ~SciPost Physics }}
\rhead{{\bf \color{scipostdeepblue} ~Submission }}

\fancyfoot[C]{\textbf{\thepage}}
}

\begin{document}

\pagestyle{SPstyle}

\begin{center}{\Large \textbf{\color{scipostdeepblue}{
Bootstrapping the $R$-matrix\\
}}}\end{center}

\begin{center}\textbf{
Zhao Zhang\textsuperscript{1$\star$}
}\end{center}

\begin{center}
{\bf 1} Department of Physics, University of Oslo, P.O. Box 1048 Blindern, Oslo, N-0316, Norway
\\[\baselineskip]
$\star$ \href{mailto:zhao.zhang@fys.uio.no}{\small zhao.zhang@fys.uio.no}
\end{center}

\section*{\color{scipostdeepblue}{Abstract}}
\textbf{\boldmath{%
A bootstrap program is presented for algebraically solving the $R$-matrix of a generic integrable quantum spin chain from its Hamiltonian. The Yang-Baxter equation contains an infinite number of seemingly independent constraints on the operator valued coefficients in the expansion of the $R$-matrices with respect to their spectral parameters, with the lowest order one being the Reshetikhin condition. These coefficients can be solved iteratively in a self consistent way using a lemma due to Kennedy, which reconstructs the $R$-matrix after an infinite number of steps. For a generic Hamiltonian, the procedure could fail at any step, making the conditions useful as an integrability test. However, at least for the most common examples, they always turn out to be satisfied whenever the lowest order condition is. It remains to be understood whether they are indeed implied by the Reshetikhin condition.
}}

\vspace{\baselineskip}

\noindent\textcolor{white!90!black}{%
\fbox{\parbox{0.975\linewidth}{%
\textcolor{white!40!black}{\begin{tabular}{lr}%
  \begin{minipage}{0.6\textwidth}%
    {\small Copyright attribution to authors. \newline
    This work is a submission to SciPost Physics. \newline
    License information to appear upon publication. \newline
    Publication information to appear upon publication.}
  \end{minipage} & \begin{minipage}{0.4\textwidth}
    {\small Received Date \newline Accepted Date \newline Published Date}%
  \end{minipage}
\end{tabular}}
}}
}

\vspace{10pt}
\noindent\rule{\textwidth}{1pt}
\tableofcontents
\noindent\rule{\textwidth}{1pt}
\vspace{10pt}

\section{Introduction}
\label{sec:intro}

In statistical mechanics courses, the $R$-matrix is usually introduced in vertex models as a way to compactly encode the Boltzmann weights, which can be used to conveniently compute the partition function with the transfer matrix method. The success of this calculation depends on the commutativity between column transfer matrices, which is guaranteed by the Yang-Baxter equation (YBE) satisfied by integrable models. For every two-dimensional (2D) integrable statistical mechanical model, one can derive the quantum Hamiltonian for a related one-dimensional integrable spin chain from the transfer matrix. However, it is not always clear how to find the 2D statistical mechanical dual model from the Hamiltonian of a quantum spin chain. 

Nevertheless, without an $R$-matrix that can be used to test integrability with the YBE, the integrability of short-range interacting quantum spin chains has been successfully tested by the Reshetikhin condition~\cite{Kulish:1982aa}, which involves only the local Hamiltonian terms. It has hence been conjectured to be a sufficient condition for quantum integrability~\cite{Grabowski:1995aa}.\footnote{In the same reference, the existence of a three-local conserved charge (not necessarily resulting from the Reshetikhin condition) was also conjectured to be a necessary condition for integrability.} While the sufficiency seems to have been established in a recent proof \cite{hokkyo2025integrabilitysingleconservationlaw}, it remains unclear whether the Reshetikhin condition implies integrability in the YBE sense. To answer this open question, an $R$-matrix satisfying the YBE has to be constructed from the integrable Hamiltonian.

In practice, the Reshetikhin condition can been checked using Kennedy's inversion formula \cite{Kennedy:1992aa}, which is actually also the first step towards constructing the $R$-matrix from an integrable Hamiltonian that has so far been overlooked. To fully unleash the power of Kennedy's method, one can try Taylor expanding a fictitious $R$-matrix with respect to its spectral parameter demanding the first order term to be proportional to the Hamiltonian in question. This has been explored in Ref.~\cite{Mutter:1995aa, BIBIKOV2003209}, in order to obtain higher order integrability tests than Reshetikhin's condition. However, their attempts were not so successful, for two good reasons. Firstly, they did not use the identities implied by the lower order terms in the YBE to simply the higher order ones to a symmetric form that resembles the Reshetikhin condition. Secondly and more importantly, they have not considered the fact that redefining the zero energy of a system should not make a difference for its integrability. Indeed, the relation between the Hamiltonian and derivative of its corresponding $R$-matrix is only up to a constant shift. This subtlety becomes important for integrable examples such as the Takhtajan-Babujian spin-1 model~\cite{Takhtajan,Babujian}, which will be elaborated in Appendix \ref{sec:spin-1}. Without the inclusion of the constant shift parameter, such integrable Hamiltonians, despite satisfying the Reshetikhin condition, could violate the second order integrability condition that Ref.~\cite{Mutter:1995aa, BIBIKOV2003209} obtained.

In this article, both considerations are taken into account to construct a bootstrap program for iteratively solving the $R$-matrix corresponding to any integrable local Hamiltonian as long as its $R$-matrix depends only on one spectral parameter. The procedure is new in that the computations are purely algebraic, without having to solve any differential equations. Therefore it can be implemented in symbolic calculations at a very affordable cost even for large local degrees of freedom. At each step of the program, a higher order version of the Reshetikhin condition is used, which are also apparently independent integrability conditions. Together, they are equivalent to the YBE of difference form, bearing close resemblance to the infinite conserved charges but depends only on three neighboring lattice sites instead. In addition, Kennedy's inversion formula is also generalized to operators with larger supports that allows the computation of explicit forms of the charge densities. The generalization itself reveals a discrete conformal algebra structure, which together with an analysis of the lattice Poincair\'e group strengthens the understanding of higher order charges as the same conserved charge observed in different reference frames.

The rest of the article is organized as follows. Sec.~\ref{sec:bootstrap} reviews the Reshetikhin condition as derived from Taylor expanding the $R$-matrix with respect to spectral parameter, and its relation to the conservation of the three-local charge. Sec.~\ref{sec:higher} derives higher order forms of the integrability conditions, highlighting the apparent independence among them, and their usage as integrability test for a generic Hamiltonian, and iterative construction of the $R$-matrix for integrable Hamiltonians. Sec.~\ref{sec:sl} parametrizes generic bi-local Hamiltonians with $\mathfrak{sl}(n,\mathbb{C})$ matrices to illustrate how the condition can be transformed into equations involving the coupling constants of the interaction. Sec.~\ref{sec:boost} introduces the boost operator that generates higher conserved charges from lower ones, and the associated time dependent symmetry manifest in the Heisenberg picture. Sec.~\ref{sec:Poincare} takes an interlude of the lattice Poincar\'e group to give an intuition of how higher conserved charges generated by the boost operator could possibly be the same three-local charge observed in different reference frames. The heuristic picture gained in Sec.~\ref{sec:Poincare} is then connected back to the theme of conserved charges by the introduction of a discrete conformal algebra in Sec.~\ref{sec:conformal}. The difficulties of generalizing the bootstrap program to non-relativistic integrable Hamiltonians with $R$-matrices that depends on two spectral parameters are explained in Sec.~\ref{sec:general}. Finally, the article concludes by discussing a few possible future directions in Sec.~\ref{sec:conclusion}.

\section{Reshetikhin's condition and the continuity equation}\label{sec:bootstrap}

The starting point of our discussion is a quantum spin chain with local interactions between nearest neighbors. The Hamiltonian is of the form $H=\sum_{x}h_{x,x+1}$, where the bi-local operator $h_{x,x+1}=\cdots\otimes 1_{x-1}\otimes h_{x,x+1}\otimes 1_{x+2}\otimes \cdots$ acts non-trivially only on the tensor product of local Hilbert space at site $x$ and $x+1$. The range of the sum in the definition of the Hamiltonian has been deliberately omitted by assuming the chain to be either infinitely long, so that we do not have to worry about boundary conditions. By definition of the local Hamiltonian operators $h_{x,x+1}$, it is apparent that they commute with each other unless their supports overlap: $[h_{x,x+1},h_{x',x'+1}]=0$ if $x'\ne x-1, x+1$.

For integrable Hamiltonians with $R$-matrices that depend only on the the difference of two spectral parameters or rapidities, a hypothetical $R$-matrix that depends on only one spectral parameter can be constructed from the Hamiltonian density $h_{x,x+1}$,
\begin{equation}
	\check{R}_{x,x+1}(\xi)=1_{x,x+1}
    +\sum_{n=1}^\infty \frac{\xi^n}{n!} \check{R}_{x,x+1}^{(n)},
	\label{eq:Rmat}
\end{equation}
where $\check{R}_{x,x+1}^{(1)}=h_{x,x+1}+c1_{x,x+1}$.\footnote{Alternatively, one can consider the expansion $\check{R}_{x,x+1}(\xi)=f(\xi)1_{x,x+1}+ \xi h_{x,x+1}+\sum_{n=2}^\infty \frac{\xi^n}{n!} \check{R}_{x,x+1}^{(n)}$.} The importance of the constant $c$ will be become clear in Sec.~\ref{sec:higher} and be exemplified in Appendix \ref{sec:spin-1}, but essentially it comes from the fact that neither an overall rescaling or a constant shift of the Hamiltonian should alter whether the system is integrable or not: as long as there is a choice of $c$ that makes \eqref{eq:Rmat} satisfy the YBE, the Hamiltonian is integrable.

The $R$-matrix so constructed from a relativistic integrable Hamiltonian satisfies the braided form of the YBE
\begin{equation}
\check{R}_{x,x+1}(\zeta)\check{R}_{x-1,x}(\xi)\check{R}_{x,x+1}(\xi-\zeta)\\ =\check{R}_{x-1,x}(\xi-\zeta)\check{R}_{x,x+1}(\xi)\check{R}_{x-1,x}(\zeta).
\label{eq:YBE}
\end{equation}
Taking $\xi=0$, it implies the unitarity condition
\begin{equation}
	\check{R}_{x,x+1}(\zeta)\check{R}_{x,x+1}(-\zeta)=1_{x,x+1}.
	\label{eq:unitarity}
\end{equation}
At even powers of $\zeta$, \eqref{eq:unitarity} fixes the even order operators in \eqref{eq:Rmat} in terms of lower order ones
\begin{equation}
	\check{R}_{x,x+1}^{(2m)}=\frac{1}{2}\sum_{k=1}^{2m-1}(-1)^{k-1}\binom{2m}{k}\check{R}_{x,x+1}^{(k)}\check{R}_{x,x+1}^{(2m-k)},
	\label{eq:unieven}
\end{equation}which makes $\check{R}_{x,x+1}^{(2)}=(h_{x,x+1}+c1_{x,x+1})^2$.\footnote{The condition from odd powers of $\zeta$
\begin{equation}
	\sum_{k=1}^{2m}(-1)^k\binom{2m+1}{k}\check{R}^{(k)}_{x,x+1}\check{R}^{(2m+1-k)}_{x,x+1}=0.
	\label{eq:uniodd}
\end{equation} are always ensured by those from the lower even powers.}
Plugging \eqref{eq:Rmat} into \eqref{eq:YBE}, and collecting the coefficients of terms proportional to $\xi\zeta^2$ gives Reshetikhin's condition~\cite{Kulish:1982aa}

\begin{equation}
\begin{split}
	[h_{x,x+1},[h_{x,x+1},h_{x-1,x}]]-&[h_{x-1,x},[h_{x-1,x},h_{x,x+1}]]\\ =\big(\check{R}_{x-1,x}^{(3)}-(h_{x-1,x}+c)^3\big)-&\big(\check{R}_{x,x+1}^{(3)}-(h_{x,x+1}+c)^3\big).
\end{split}    
	\label{eq:Reshetikhin}
\end{equation}
Up to additional terms that disappear when summed over $x$, the LHS is the commutator between the Hamiltonian and the charge density $\rho_{x}\equiv [h_{x,x+1},h_{x-1,x}]$, while the RHS is a divergence. So \eqref{eq:Reshetikhin} is nothing but the continuity equation
\begin{equation}
	\partial_t\rho_{x}=i[H,\rho_{x}]=j_{x}-j_{x+1}
	\label{eq:continuity}
\end{equation}where the unit $\hbar=1$ has been adopted.
Performing a spatial sum, the RHS cancels telescopically and one recovers the conservation of the three-local charge $Q= \sum_x \rho_{x}$:
\begin{equation}
    [H,Q]=\sum_{x=1}^L\Big([h_{x-1,x},[h_{x+1,x+2},h_{x,x+1}]] +[H,\rho_{x}] +[h_{x+1,x+2},[h_{x,x+1},h_{x-1,x}]] \Big)=0,
\label{eq:Q3}
\end{equation}where the first and last term in the summand, accounting for the difference between the LHSs of \eqref{eq:Reshetikhin} and \eqref{eq:continuity}, cancel due to the Jacobi identity.

\section{The $R$-matrix bootstrap} \label{sec:higher}

Higher order generalizations to Reshetikhin's condition \eqref{eq:Reshetikhin} and the yet to be defined operator $\check{R}_{j,j+1}^{(2m+1)}$ can be obtained from any of the coefficients of terms homogeneous to $\xi^p\zeta^{2m+1-p}$, for $1\le p\le2m$. Out of  the $2m$ relations, only one is independent. After a manipulation detailed in Appendix \ref{sec:flux}, the identities arising from terms proportional to $\xi\zeta^{2m}$ become
\begin{equation}
\begin{split}
    &\frac{1}{2}\sum_{k=1}^{2m-1}(-1)^k\binom{2m}{k}\left(\left[\check{R}^{(k)}_{12},\left[\check{R}^{(1)}_{23},\check{R}^{(2m-k)}_{12}\right]\right]-\left[\check{R}^{(k)}_{23},\left[\check{R}^{(1)}_{12},\check{R}^{(2m-k)}_{23}\right]\right]\right) \\ =&\sum_{k=1}^{2m+1}(-1)^k\binom{2m}{k-1}\left(\check{R}^{(k)}_{12}\check{R}^{(2m+1-k)}_{12}-\check{R}^{(k)}_{23}\check{R}^{(2m+1-k)}_{23}\right).
\end{split}
\label{eq:highercurrent}
\end{equation} Notice that an independent condition appears at every second order in the expansion. This should be compared with the algebraically independent charge conservation laws to be discussed in the next section (see the right panel of Fig.~\ref{fig:squid}). Despite the similarity to \eqref{eq:Reshetikhin}, it is not clear this time whether \eqref{eq:highercurrent} implies the charge conservation of $[H,Q^{(2m+1)}]=0$ in a similar way as shown by \eqref{eq:Q3}. It is therefore an interesting open question whether a direct relation can be established between the two sets of conditions, such that \eqref{eq:highercurrent} acting on three neighboring sites for all $m$ implies $[H,Q^{(2m+1)}]=0$, which depends on $2m+1$ sites, or even vice versa. Furthermore, in light of the result of Ref.~\cite{hokkyo2025integrabilitysingleconservationlaw} and lacking a counterexample, the independence of \eqref{eq:highercurrent} from the Reshetikhin condition could also be questioned. Nevertheless, once the $R$-matrix is constructed order by order, obtaining the conserved charges is simply a matter of expanding the monodromy or column transfer matrix, the commutativity among which at different spectral parameters is guaranteed by the YBE. 

Before detailing this standard procedure in the Sec.~\ref{sec:boost}, I first explain how Eq.~\eqref{eq:highercurrent} can be used to solve $\check{R}^{(2m-1)}_{x,x+1}$ from only the Hamiltonian $h_{x,x+1}$. Together with \eqref{eq:unieven}, this serves as a way to bootstrap the $R$-matrix $\check{R}(\xi)$ for fundamental integrable Hamiltonians. Kennedy came up with a way to find $\check{R}^{(3)}_{x,x+1}$ from \eqref{eq:Reshetikhin} (Lemma 1 in Ref.~\cite{Kennedy:1992aa}), which he used to find integrable SU(2) isotropic spin chains. Now, suppose we know already all $\check{R}_{12}^{(k)}$ for $k\le 2m$, then the LHS of \eqref{eq:highercurrent} is completely known, and so is the RHS except the term for $k=2m+1$, i.e.~$\check{R}_{23}^{(2m+1)}-\check{R}_{12}^{(2m+1)}$. So the lemma by Kennedy can be applied again to find the RHS, which give these two unknowns. 

Although by brute force it seems that the $R$-matrix can only be fully constructed in this procedure after an infinite number of steps, drastic shortcuts can be made for smaller local Hilbert space dimensions. One example is to take advantage of the Cayley-Hamiltonian theorem to reduce the number of independent parameters, as used in Ref.~\cite{deLeeuw_2019} for the XYZ chain.\footnote{The infinite number of elementary steps can also be traded for one step of solving Sutherland's equation to find the $R$-matrix~\cite{PhysRevLett.125.031604,10.21468/SciPostPhys.11.3.069}. This approach works also for non-relativistic $R$-matrices, and has been used to classify integrable models with low-dimensional local Hilbert space~\cite{corcoran2024regular4times4,deleeuw20254x4solutions}.} Another dream scenario is that after just a few iterations, we can make an educated guess of the matrix elements as functions of the spectral parameter from the first few terms of their Taylor expansions. When this happens, there is also hope to define an integrable 2D statistical mechanical with the Boltzmann weights for classical configurations determined from the $R$-matrix elements, in the same way that XXZ spin chain is related to the six-vertex model. This could be a future direction worth exploring for those quantum integrable models that do not currently have a known classical counterpart.

The presentation so far has been assuming a Hamiltonian to be integrable, from which identities like \eqref{eq:Reshetikhin} and \eqref{eq:highercurrent} are derived as consequences. Now suppose instead we are given a Hamiltonian and our goal is to find out whether it is integrable or not. For instance, to check Reshetikhin's condition \eqref{eq:Reshetikhin}, first construct the candidate two-local operator
\begin{equation}
    \tilde{j}_x=\tr_{x+1}[h_{x-1,x}+h_{x,x+1},\rho_{x}]+\tr_{x+1,x+2}[h_{x,x+1}+h_{x+1,x+2},\rho_{x+1}]
	\label{eq:Kennedy}
\end{equation}
where $\tr_x$ is the partial trace over the local Hilbert space at site $x$ divided by its dimensionality, such that $\tr_x 1_x=1$. Again, this `surface flux' $\tilde{j}_x$ (as a function of $c$) can be calculated just as well for non-integrable Hamiltonians. So the real test is to check if the identity
\begin{equation}
    [h_{x-1,x}+h_{x,x+1},\rho_{x}]=\tilde{j}_{x}\otimes1_{x+1}- 1_{x-1}\otimes \tilde{j}_{x+1}
    \label{eq:Kennedytest}
\end{equation}holds. If the LHS indeed turns out to be a pure divergence, since the RHS of \eqref{eq:Reshetikhin} involves only known operators except $\check{R}_{x,x+1}^{(3)}$, it can be solved from $\tilde{j}_{x}$ (as a function of $c$). Notice that different choices of $c$ naturally result in different $\check{R}_{x,x+1}^{(3)}$, which would affect the LHS of \eqref{eq:highercurrent} at $m=2$. Suppose for $m=2$ and $x=2$, \eqref{eq:highercurrent} is given by $-E_{123}(c)=Y_{12}(c)-Y_{23}(c)$, then Kennedy's lemma gives a candidate $\tilde{Y}_{12}(c)=-\tr_3E_{123}(c)-\tr_{34}E_{234}(c)$. The condition at $m=2$ therefore amounts to $-E_{123}(c)=\tilde{Y}_{12}(c)-\tilde{Y}_{23}(c)$, which gives a set of equations on the unknown $c$ when considered element-wise. As long as there is a solution for $c$, the additional condition is considered satisfied up to this step, and one can carry on to higher order conditions fixing $c$ to the solution. On the other hand, if there is no solution, the outcome of the integrability test is conclusively negative.

After understanding the role of the constant $c$, we can modify the above procedure to find integrable Hamiltonians. We start with a parametrization of generic Hamiltonians subject to certain symmetry, such that each matrix element of $h_{x,x+1}$ is a multivariate function. Then the above procedure at each step would result in a set of equations on these parameters. Solving \eqref{eq:Kennedytest} as equations on these unknown variables tells us which particular choices in the parameter space are integrable.

While it has proved to be a very affordable symbolic computation using \texttt{Mathematica}, it seems that the criteria need to be checked to infinite order to be sure that the Hamiltonian is integrable in theory. Surprisingly, no known example of non-integrable Hamiltonians fails the test at any later stage than the first step. Put differently, it seems that at least empirically it is redundant to check \eqref{eq:highercurrent} once the Reshetikhin condition is satisfied. So the moral of the story is that while the conservation of the higher order charges to be discussed in Sec.~\ref{sec:boost} does not automatically guarantee the satisfaction of higher order Reshetikhin conditions, the possibility of a more intricate connection allowing the Reshetikhin condition to guarantee its higher order counterparts is not excluded. Moreover, even if the higher order Reshetikhin conditions turn out to be equivalent to the conservation of higher charges generated by the boost operator, which according to Ref.~\cite{hokkyo2025integrabilitysingleconservationlaw} are implied by the Reshetikhin condition, the explicit form of the $R$-matrix has yet to be found in some constructive manner. The bootstrap program provides a foolproof solution to this problem.

\section{$\mathfrak{sl}(n,\mathbb{C})$ Hamiltonians}\label{sec:sl}

In this section, the formal description of the bootstrap program above is concretized by parameterizing the Hamiltonian in terms of $\mathfrak{sl}(n,\mathbb{C})$ matrices. In this way, we can arrive at an explicit expression of $\check{R}_{12}^{(3)}$ for a generic $h_{12}$. The later steps of program are more suitably implemented in automated symbolic calculation, as illustrated by the \texttt{Mathematica} code provided in the supplementary information. 

For a local Hilbert space with $n$ dimensions, any local operator $M$ can be expressed as 
\begin{equation}
    M=\sum_{\alpha=1}^{n^2-1}m_\alpha T^{\alpha}+m_0 1_{n\times n},
\end{equation}where the matrices $T^{\alpha}, \alpha=1,\cdots n^2-1$ form a basis of traceless $n\times n$ matrices. Since $\mathfrak{sl}(n,\mathbb{C})$ is the complexification of $\mathfrak{su}(n)$, it is more convenient to let the coefficients $m_\alpha$ be complex numbers, so that the generators $T^{\alpha}$ become Hermitian matrices forming a representation of $\mathfrak{su}(n)$ and satisfying the commutation and anti-commutation relations
\begin{equation}
    [T^\alpha,T^\beta]= if_\gamma^{\alpha\beta} T^\gamma, \quad \{T^\alpha,T^\beta\}= 2\delta^{\alpha\beta}1_{n\times n}+2d_\gamma^{\alpha\beta} T^\gamma, \label{eq:commutation}
\end{equation}with the Kronecker delta $\delta^{\alpha\beta}=1$ if $\alpha =\beta$; and 0 if $\alpha \ne\beta$. $d_\gamma^{\alpha\beta}$ is totally symmetric about all three indices, while the structure constants $f_\gamma^{\alpha\beta}$ is totally antisymmetric, and satisfy the Jacobi identity\footnote{Furthermore, the graded Jacobi identities
\begin{equation}
\begin{split}
    [\{T^\alpha,T^\beta\},T^\gamma]+[\{T^\beta,T^\gamma\},T^\alpha]+[\{T^\gamma,T^\alpha\},T^\beta]=&0,\\
    [\{T^\alpha,T^\beta\},T^\gamma]+\{[T^\gamma,T^\beta],T^\alpha\}+\{[T^\gamma,T^\alpha],T^\beta\}=&0
\end{split}
\end{equation}gives the properties (which has not been used though in the following derivations)
\begin{equation}
\begin{split}
        d_\delta^{\alpha\beta}f_\epsilon^{\delta\gamma}+d_\delta^{\beta\gamma}f_\epsilon^{\delta\alpha}+d_\delta^{\gamma\alpha}f_\epsilon^{\delta\beta}=&0\\
        d_\delta^{\alpha\beta}f_\epsilon^{\delta\gamma}+f_\delta^{\gamma\beta}d_\epsilon^{\delta\alpha}+f_\delta^{\gamma\alpha}d_\epsilon^{\delta\beta}=&0. 
\end{split}
\label{eq:Jacobi2}
\end{equation}} 
\begin{equation}                                
    f_\delta^{\alpha\beta}f_\epsilon^{\delta\gamma}+f_\delta^{\beta\gamma}f_\epsilon^{\delta\alpha}+f_\delta^{\gamma\alpha}f_\epsilon^{\delta\beta}=0.\label{eq:Jacobi1}
\end{equation}

Thus any bi-local Hamiltonian operator can be written as
\begin{equation}
    h_{j,j+1}=\tilde{a}_{\alpha\beta}\tilde{T}^\alpha_j \tilde{T}^\beta_{j+1}+ \tilde{b}_\alpha (\tilde{T}_j^\alpha + \tilde{T}_{j+1}^\alpha)+c, \label{eq:Hamiltonian}
\end{equation}where from now on summation over dummy Greek indices are implied. Parity symmetry requires also that $\tilde{a}_{\beta\alpha}=\tilde{a}_{\alpha\beta}$.\footnote{Notice that this puts no constraint on the Hermiticity of the Hamiltonian $h_{j,j+1}$, since $T^{\alpha}$ are chosen to be Hermitian themselves. Non-Hermitian Hamiltonians instead have $a^*_{\alpha\beta}\ne a_{\alpha\beta}$ or $b^*_{\alpha}\ne b_{\alpha}$, which is not excluded from the following discussions.} This means we can change to the basis in which the matrix of coefficients become diagonal. Since linear superpositions of traceless Hermitian matrices remain traceless and Hermitian, the structure \eqref{eq:commutation} still holds, with modified constants. Hence in the diagonal basis, the Hamiltonian \eqref{eq:Hamiltonian} can be written as
\begin{equation}
    h_{j,j+1}=a_{\alpha}T^\alpha_j T^\alpha_{j+1}+ b_\alpha (T_j^\alpha + T_{j+1}^\alpha)+c. \label{eq:sunham}
\end{equation}
The commutator between neighboring Hamiltonians can be written as
\begin{equation}
    [h_{12},h_{23}]=ia_{\alpha}a_{\gamma}f_\beta^{\alpha\gamma}T^\alpha_1 T^\beta_{2} T^\gamma_3+ia_{\alpha }b_\gamma f_\beta^{\alpha\gamma}T^\alpha_1 T^\beta_{2}+ia_{\beta}b_\gamma f_\alpha^{\gamma\beta}T^\alpha_{2} T^\beta_3 .
\end{equation}
To get the next commutators, we need
\begin{equation}
    \begin{split}
        [T_j^\alpha\otimes T_{j+1}^\beta,T_j^\gamma\otimes T_{j+1}^\delta]=&\frac{1}{2}\left([T_j^\alpha,T_j^\gamma]\otimes\{T_{j+1}^\beta,T_{j+1}^\delta\}+\{T_j^\alpha,T_j^\gamma\}\otimes[T_{j+1}^\beta,T_{j+1}^\delta]\right)\\
	=&i\left(f^{\alpha\gamma}_\epsilon d_\zeta^{\beta\delta}+f^{\beta\delta}_{\zeta}d^{\alpha\gamma}_{\epsilon}\right)T_j^{\epsilon}T^\zeta_{j+1}+if^{\alpha\gamma}_\epsilon \delta^{\beta\delta}T_j^\epsilon+if^{\beta\delta}_{\epsilon}\delta^{\alpha\gamma}T_{j+1}^{\epsilon}.
    \end{split}\label{eq:tensorcommutator}
\end{equation}
Tedious but elementary derivations show that terms on the LHS of \eqref{eq:Reshetikhin} can be put in the form
\begin{equation}
\begin{split}
    [h_{12},[h_{12},h_{23}]]=&u_{\alpha\beta\gamma}T_1^\alpha T_2^\beta T_3^\gamma+v_{\alpha\beta}T_1^\alpha T_2^\beta+w_{\alpha\beta}T_1^\alpha T_3^\beta+x_{\alpha\beta}T_2^\alpha T_3^\beta+y_\alpha T_1^\alpha+z_\alpha T_2^\alpha,\\
    [h_{23},[h_{23},h_{12}]]=&u_{\gamma\beta\alpha}T_1^\alpha T_2^\beta T_3^\gamma+x_{\beta\alpha}T_1^\alpha T_2^\beta+w_{\alpha\beta}T_1^\alpha T_3^\beta+v_{\beta\alpha}T_2^\alpha T_3^\beta+z_\alpha T_2^\alpha+y_\alpha T_3^\alpha,
\end{split}\label{eq:LHS}
\end{equation}with the coefficients after simplifications using the symmetry properties of the indices and \eqref{eq:Jacobi1} becoming
\begin{equation}
    \begin{split}
        u_{\alpha\beta\gamma}=&a_{\gamma}a_{\delta}a_{\epsilon}f_\zeta^{\gamma\delta}(f_\alpha^{\epsilon\delta}d_\beta^{\epsilon\zeta}+f_\beta^{\epsilon\zeta}d_\alpha^{\epsilon\delta})+a_{\gamma}b_\delta(a_{\alpha}f_\gamma^{\delta\epsilon}f_\beta^{\alpha\epsilon}+a_{\epsilon}f_\beta^{\gamma\epsilon}f_\alpha^{\delta\epsilon}),\\
        v_{\alpha\beta}=&a_{\gamma}a_{\delta}b_\epsilon f_\zeta^{\epsilon\delta}(f_\alpha^{\gamma\delta}d_\beta^{\gamma\zeta}+f_\beta^{\gamma\zeta}d_\alpha^{\gamma\delta})+b_\gamma b_\delta (a_{\epsilon}f_\beta^{\gamma\epsilon}f_\alpha^{\delta\epsilon}-a_{\alpha}f_\alpha^{\gamma\epsilon}f_\beta^{\delta\epsilon}),\\
        w_{\alpha\beta}=&a_{\beta}a_{\gamma}a_{\delta}f_\alpha^{\gamma\delta}f_\beta^{\delta\gamma},\\
        x_{\alpha\beta}=&a_{\beta}b_\gamma b_\delta f_\alpha^{\gamma\epsilon}f_\beta^{\delta\epsilon},\\
        y_\alpha=& a_{\beta}a_{\gamma}b_\delta f_\alpha^{\beta\gamma}f_\delta^{\gamma\beta},\\ 
        z_\alpha=& a_{\beta}a_{\beta}b_\gamma f_\alpha^{\beta\delta}f_\gamma^{\beta\delta}.
    \end{split}
\end{equation}
Now using Kennedy's trace trick and the traceless property of $T^{\alpha}_j$, we have 
\begin{equation}
\begin{split}
       X_{12}=&-\tr_3D_{123}-\tr_{34}D_{234}
       =(x_{\beta\alpha}-v_{\alpha\beta})T_1^\alpha T_2^\beta-y_\alpha (T_1^\alpha+ T_2^\alpha),\\
       X_{23}=&\tr_1 D_{123}+\tr_{01}D_{012}=(x_{\alpha\beta}-v_{\beta\alpha})T_2^\alpha T_3^\beta-y_\alpha( T_2^\alpha+ T_3^\alpha)
\end{split}
\end{equation}In order for them to be the same operator acting on different degrees of freedom, we must have $v_{\alpha\beta}+x_{\alpha\beta}=v_{\beta\alpha}+x_{\beta\alpha}$, $\forall \alpha,\beta$. In addition, $-D_{123}=X_{12}-X_{23}$ requires $u_{\alpha\beta\gamma}=u_{\gamma\beta\alpha}$, $\forall \alpha,\beta,\gamma$. Explicitly, they are written as
\begin{equation}
    \begin{split}
         \left(f^{\epsilon\delta}_\alpha d^{\epsilon\gamma}_\beta-f^{\epsilon\delta}_\beta d^{\epsilon\gamma}_\alpha\right)f^{\gamma\delta}_\zeta\left(a_\gamma+a_\delta\right)a_\epsilon b_\zeta   &=0, \quad \forall \alpha \ne \beta,\\
         a_\delta a_\epsilon \left[d^{\delta\zeta}_\beta\left(a_{\gamma}f^{\gamma\epsilon}_\zeta f^{\alpha\epsilon}_\delta-a_{\alpha}f^{\alpha\epsilon}_\zeta f^{\gamma\epsilon}_\delta\right)+f^{\delta\zeta}_\beta\left(a_\gamma f^{\epsilon\gamma}_\zeta d^{\delta\epsilon}_\alpha - a_{\alpha}f_\zeta^{\epsilon\alpha}d^{\delta\epsilon}_\gamma\right)\right]\\ +a_\epsilon b_\delta\left(a_\gamma f^{\epsilon\gamma}_\beta f^{\delta\epsilon}_\alpha-a_\alpha f^{\epsilon\alpha}_\beta f^{\delta\epsilon}_\gamma\right)&=0, \quad \forall \beta,\ \mathrm{and}\ \alpha\ne\gamma.
    \end{split}\label{eq:explicitcondition}
\end{equation}

To understand these conditions, let us look at an example of $n=2$. In this case, $d^{\alpha\beta}_\gamma=0$,
so the only condition left is $a_\epsilon b_\delta\left(a_\gamma f^{\epsilon\gamma}_\beta f^{\delta\epsilon}_\alpha-a_\alpha f^{\alpha\epsilon}_\beta f^{\epsilon\delta}_\gamma\right)=0$. Due to the total antisymmetry of the structure constant, and between the indices $\alpha$ and $\gamma$, there are six conditions that have to be satisfied simultaneously
\begin{equation}
    a_1a_2b_1=a_1a_2b_2=a_1a_3b_1=a_1a_3b_3=a_2a_3b_2=a_2a_3b_3=0.
\end{equation}The solutions can be classified as follows: In the case $b_1=b_2=b_3=0$, there is no constraint on $a_{1,2,3}$, meaning that XYZ models are integrable. In the case $b_1=b_2=0$, we need further either $a_1=a_2=0$, giving the longitudinal field Ising model, or $a_3=0$, known as the XYh model. The case $b_3=0$ has to be accompanied by $a_1=a_2=0$, which is the transverse field Ising model. Finally, if all three linear terms are present, the Hamiltonian can only be integrable if it is non-interacting, that is, $a_1=a_2=a_3=0$. These are all of the integrable models with two local degrees of freedom, as have been well classified using the conserved charge approach \cite{yamaguchi2024complete,yamaguchi2024proof}. 

The full classification of integrable models with larger local Hilbert space is left for future works, but a simple observations can be made. If there is no linear terms in the Hamiltonian, the condition becomes
\begin{equation}
    a_\delta a_\epsilon \left[d^{\delta\zeta}_\beta\left(a_{\gamma}f^{\epsilon\gamma}_\zeta f^{\delta\epsilon}_\alpha-a_{\alpha}f^{\alpha\epsilon}_\zeta f^{\epsilon\delta}_\gamma\right)+f^{\delta\zeta}_\beta\left(a_\gamma f^{\epsilon\gamma}_\zeta d^{\delta\epsilon}_\alpha - a_{\alpha}f_\zeta^{\epsilon\alpha}d^{\delta\epsilon}_\gamma\right)\right]=0,
\end{equation}which is a set of equations on the coefficients $a_\alpha$. So anisotropic higher spin chains and Hamiltonians explicitly breaking SU($N$) symmetry are generically non-integrable, as has been studied in Ref.~\cite{Kennedy:1992aa, PhysRevB.106.134420}.

In Appendix \ref{sec:models}, the construction in this section will be used to compute the explicit expression for $\check{R}_{12}^{(3)}=X_{12}+h_{12}^3$ for a few examples of integrable models.

\section{Conserved charges and the boost operator}\label{sec:boost}

In recent decades, quantum integrability has been much more heavily investigated in terms of symmetry currents and conserved charges than in the framework of YBE \cite{Caux_2011}. So this section is dedicated to making a connection between the two contexts highlighting the fact that the YBE is a stronger condition than existence of infinite conserved charges.

From the $R$-matrix obtained in the previous section, one can construct the monodromy matrix $\mathfrak{T}(\xi)=\prod_{x} R_{x,x+1}(\xi)$,\footnote{For a finite chain with periodic boundary condition, the trace can to be taken to get the column transfer matrix.} where $R_{x,x+1}(\xi)=P_{x,x+1}\check{R}_{x,x+1}(\xi)$, with $P_{x,x+1}$ being the permutation between site $x$ and $x+1$. It can be checked that the YBE in the form 
\begin{equation}
R_{x,x+1}(\zeta)R_{x-1,x+1}(\xi)R_{x-1,x}(\xi-\zeta) =R_{x-1,x}(\xi-\zeta)R_{x-1,x+1}(\xi)R_{x,x+1}(\zeta).
\label{eq:YBEa}
\end{equation} ensures the commutativity between monodromy operators with different spectral parameters $[\mathfrak{T}(\xi),\mathfrak{T}(\zeta)]=0$. Therefore, the charges defined by
\begin{equation*}
     \ln \mathfrak{T}(\xi)=\sum_{n=0}^\infty \frac{\xi^n}{n!}Q^{(n+1)},\ \mathrm{or}\quad Q^{(n+1)}=\frac{d^{n}}{d\xi^{n}}\ln\mathfrak{T}(\xi)\Bigr|_{\xi=0}
\end{equation*} all commute $[Q^{(m)},Q^{(n)}]=0$, with the first three being $Q^{(1)}=P\equiv\sum_x\ln P_{x,x+1}$, $Q^{(2)}=H$ and $Q^{(3)}=Q$. This way the explicit forms of the higher order conserved charges can always be obtained from the expansion of the vertex operator when the YBE is satisfied, even though their densities are not readily available.\footnote{This is because the support of the densities do not grows by one site instead of two from order to order due to similar cancellations as in \eqref{eq:Q3}.}

A more convenient way to obtain the explicit form of charges is via the boost operator~\cite{1982JETP}, defined as $ B=\sum_x xh_{x-1,x}$. The name is fitting for its operation on the monodromy matrix, if its spectral parameter is interpreted as a rapidity $[B, \mathfrak{T}(\xi)]=\partial_\xi\mathfrak{T}(\xi)$.
It follows that the boost operator acts as a ladder operator on the infinite sequence of conserved charges 
\begin{equation}
    [B, Q^{(n)}]= Q^{(n+1)}, \quad n=1,2,3,\cdots.
    \label{eq:ladder}
\end{equation}
To reveal the analogy with the generator of the Lorentz boost $K(t)=\int dx\big(xH(x)-tP(x)\big)$ in field theory, one can look at the boost operator in the Heisenberg picture, which is referred to as a time-dependent symmetry generator in the mathematics literature~\cite{timedependent}
\begin{equation}
    B(t)=e^{itH}Be^{-itH}=\sum_{n=0}^\infty\frac{(it)^n}{n!}\ad_H^n B\label{eq:timedependentsymmetry}
\end{equation}evaluated at $t=0$. If the first non-trivial charge $Q=[B,H]$ generated by $B$ is already conserved, i.e.~$[H,[H,B]]=0$, the infinite sum terminates, and the above definition becomes
\begin{equation}
    B(t)=B-itQ=\sum_x\big(xh_{x-1,x}-it\rho_{x}\big).
\end{equation}
Notice that the interpretation of the analogy to Lorentz boost is different from that given in Ref.~\cite{ThackerLorentz}, which focused instead on the boost between momentum and energy. By definition of the boost operator, the relation $[B,P]=H$ always holds, regardless of the integrability of the Hamiltonian. But as shown here, $B(t)$ only becomes a time-dependent symmetry if the infinite formal sum \eqref{eq:timedependentsymmetry} is meaningful, making it a master symmetry, when the three-local charge generated by $B$ is conserved. The interpretation of the boost operator serves as a first hint that there is a discrete Lorentz symmetry whenever Reshetikhin's condition holds, and the existence of all the higher conserved charges are automatically ensured by this symmetry. In Sec.~\ref{sec:Poincare}, we will take a closer look at how such Lorentz invariance works out when combined with lattice translation into a discrete Poincar\'e group.

\begin{figure}
	\centering
	\includegraphics[width=0.6\linewidth]{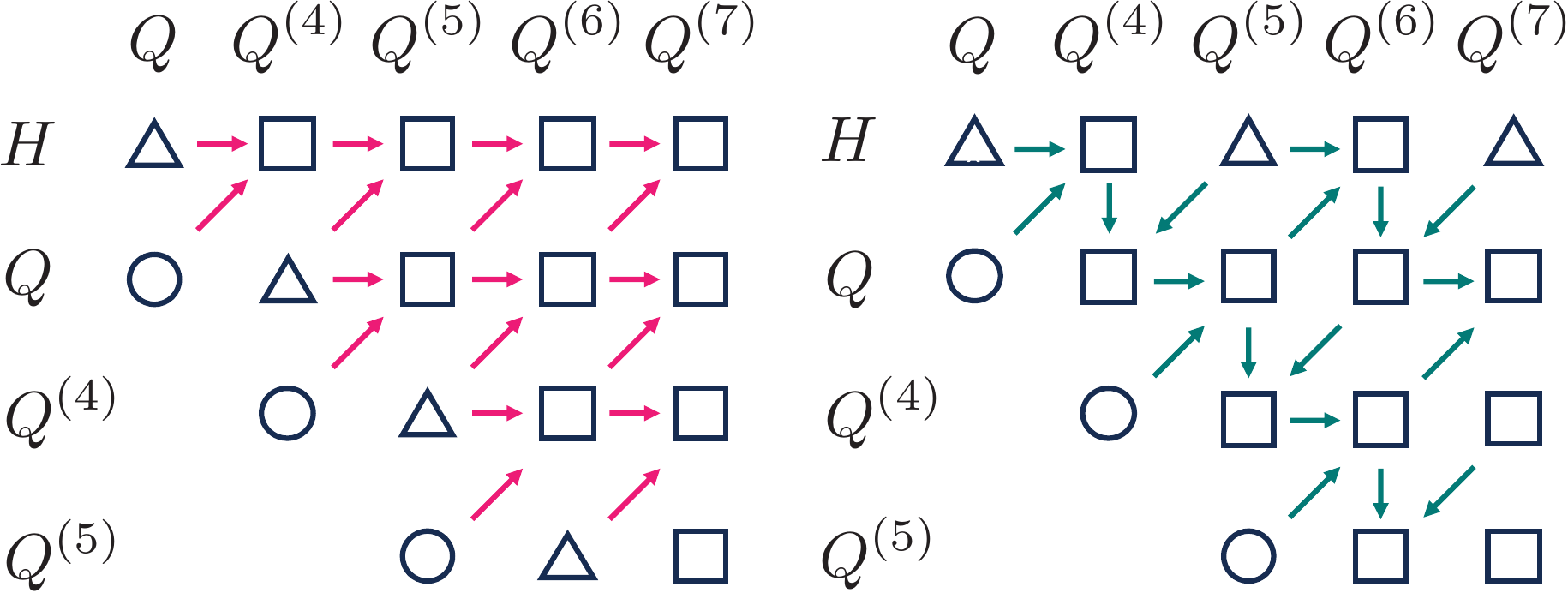}
	\caption{Two alternative schematics of the mutual commutativity of conserved charges: Circles represent the commutation between a charge with itself, which are trivially satisfied; triangles represent the independent additional conditions that imply all the other commutativity represented by squares.}
	\label{fig:squid}
\end{figure}

Unlike the definitions of classical integrability, which accommodate a spectrum of different degrees of integrability, such as integrable by quadrature, requiring the first integrals to form a closed Lie subalgebra instead of involutive, quantum integrability is usually defined by the mutual commutativity of all conserved charges, not the least due to the closely-knit YBE and transfer matrix formalism. Because of the ladder property of the boost operator, all of these commutativity cannot be independent due to the Jacobi identity
\begin{equation}
    [Q^{(m+1)},Q^{(n)}]
    =[Q^{(n+1)},Q^{(m)}]+[B,[Q^{(m)},Q^{(n)}]].\label{eq:squid}
\end{equation} Hence, \textit{at most} an $\mathcal{O}(N)$ number of the total $\mathcal{O}(N^2)$ commutativity can be independent for $N$ conserved charges.\footnote{This has of course been well known \cite{WangJP,Mariusquid}, and Ref.~\cite{hokkyo2025integrabilitysingleconservationlaw} further points out that even the algebraically independent ones are all implied by $[H, Q]=0$.} Two alternative choices of them, either $[Q^{(n)},Q^{(n+1)}]=0$ for all $n\ge 2$, or $[H,Q^{(2m+1)}]=0$ for all $m\ge 1$, are summarized in Fig.~\ref{fig:squid}. 

As observed earlier, \eqref{eq:highercurrent} could to be in one-to-one correspondence with the independent inputs of the mutual commutativity of all the conserved charges in Fig.~\ref{fig:squid}, establishing which would indicate that integrability by YBE might not be so different from its definition by infinite conserved charges. Since the conservation of the lowest order charge $Q$ is also equivalent to the Reshetikhin condition, it is no surprise that in all known examples all the higher order commutation relations are automatically satisfied when $[H,Q]=0$. However, the highly mathematical proof in Ref.~\cite{hokkyo2025integrabilitysingleconservationlaw} does not by itself lend much insight into why this works, so the next section will try to explain the physics behind based on an earlier idea by Thacker \cite{ThackerLorentz}.

\section{The lattice Poincar\'e group}\label{sec:Poincare}

Thacker argued that the algebraic structure of the conserved charges \eqref{eq:ladder} is the infinite dimensional lattice analog of the Poincar\'e algebra~\cite{ThackerLorentz}
\begin{equation}
    [P,H]=0,\quad [K,P]=iH,\quad [K,H]=iP,
    \label{eq:latLorentz}
\end{equation} where in the continuous limit the odd (resp.~even) order charges converge to $P$ (resp.~$H$). As appealing as it is, the analogy is a bit vague: While the Poincar\'e symmetry is a kinetic or spacetime symmetry, the conserved charges of integrable spin chains generate a dynamic symmetry in the internal degree of freedoms in the Hilbert space. More direct analogies were later studied as $\kappa$-deformed Poincar\'e algebras~\cite{qPoincaire,PhysRevLett.68.3718}, where the deformation parameter is related to the lattice spacing, and instead of an infinite set of generators, the group is generated by the enveloping algebra of the three operators that generate the continuous group \eqref{eq:latLorentz}~\cite{2dqPoincaire}. 

Following Ref.~\cite{Pei}, I demonstrate now how the infinite algebraic structure of \eqref{eq:ladder} naturally arises when (1+1)D Lorentz invariance is combined with discrete translation invariance. Since at least the translation in one direction is already a discrete subgroup, it is better to work with group elements instead of the Lie algebra that generates a continuous group. As a semi-direct product, the Poincar\'e group has the multiplication rule
\begin{equation}
    \big(\Lambda(\eta), \vec{\alpha} \big)\cdot \big(\Lambda(\theta), \vec{\beta} \big)=\big(\Lambda(\eta+\theta),  \vec{\alpha} +\Lambda(\eta)\vec{\beta} \big),
\end{equation}where the Lorentz boost has the two-dimensional representation
\begin{equation}
    \Lambda(\eta)=\begin{pmatrix}
    \cosh \eta & \sinh\eta\\
     \sinh\eta & \cosh \eta
    \end{pmatrix}.
\end{equation}

The (1+1)D lattice Poincar\'e group is generated by a finite translation $t_1=\big(\Lambda(0), (1\ 0)^T\big)$ and a finite boost $b(\eta)=\big(\Lambda(\eta), (0\ 0)^T\big)$. The group commutator between the two produces a translation in another direction
\begin{equation}
    t_2=[b(\eta),t_1]=b(\eta)t_1b(\eta)^{-1}t_1^{-1}=\big(\Lambda(0), \begin{pmatrix}\cosh\eta-1\\ \sinh\eta\end{pmatrix}\big).
\end{equation}
We are free to choose the new direction as the second basis vector that together span the (1+1)D spacetime. But the next commutator
\begin{equation}
    t_3=[b(\eta),t_2]=b(\eta)t_2b(\eta)^{-1}t_2^{-1}=(t_1t_2)^{2(\cosh\eta-1)}
\end{equation}has to end up on a lattice point, as required by the group closure. This only happens if $\cosh\eta$ take positive half-integer values.
$\cosh\eta=1$ is already in the group as the identity element. Among the rest of the possibilities, we can only pick one, as the product of two different boosts would end up outside the lattice. The choice of $\eta$ labels the irreducible representation.
\begin{figure}
	\centering
	\includegraphics[width=0.4\linewidth]{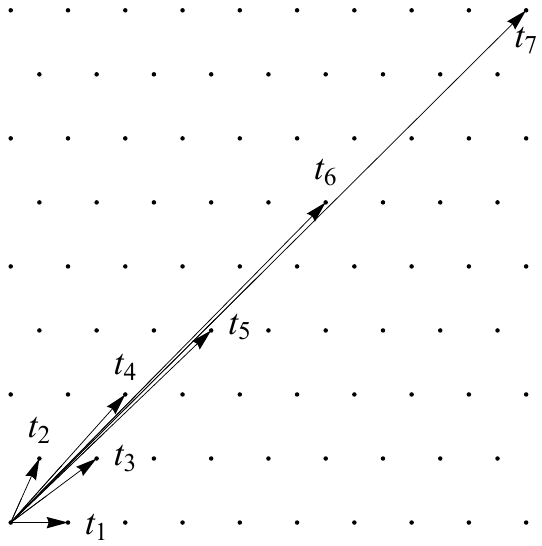}
	\caption{The lattice vectors generated by boosting the translation $t_1$ with the minimal rapidity $\arcosh\frac{3}{2}$. They alternate between space-like and time-like vectors and approach the light-cone in the infinite limit.}
	\label{fig:lattice}
\end{figure}

Now that the action of $t_3$ remains on the lattice, it follows by induction that all higher group commutators $t_{k+1}=[b(\eta),t_k]$ land on the lattice as well. In particular, for $\cosh\eta=\frac{3}{2}$, we have $t_{k+1}=t_{k}t_{k-1}$, and the corresponding lattice vectors form a Fibonacci sequence, as depicted in \cref{fig:lattice}. By analogy, the conservation of the higher conserved charges encountered in Sec.~\ref{sec:boost} could be understood as the consequence of a discrete Poincar\'e symmetry, which depends on the conservation of the three-local charge alone.

\section{The discrete conformal algebra} \label{sec:conformal}

In order to move from the lattice Poincar\'e group of spacetime coordinates to a discrete Poincar\'e algebra for conserved charges in quantum integrability, we need to find the generator of discrete translation, in a way that is compatible with the algebra among infinitesimal generators of a Lie group.\footnote{This has been done in the context of single-body quantum mechanical integrable systems~\cite{latticeQM}, and the proper mathematical tool is the umbral calculus~\cite{jordan1965calculus,boole1872treatise,PhysRevD.69.105011, umbral,Levy}.} This can be done by introducing the superoperator
\begin{equation}
    \mathcal{P}[j_x^{(n)}]=(1-\mathcal{T})[j_x^{(n)}]=j_{x}^{(n)}- j_{x+1}^{(n)},
\end{equation}
where $\mathcal{T}[j_x]=\Ad_Tj_x=Tj_xT^{-1}$ shifts the support of an operator by one lattice spacing.
A formal inverse of the difference operator can be introduced to solve the symmetry currents from the continuity equation 
\begin{equation}
    \frac{d\rho^{(n)}_x}{dt}=i[H,\rho^{(n)}_x]=j_x^{(n)}-j_{x+1}^{(n)}
\end{equation}satisfied by the higher charge densities.
Say the $n$-local density operator $\rho^{(n)}_x$ is supported by lattice sites from $x-n+2$ to $x+1$, and the $(n+1)$-local current operator $j^{(n)}_x$ acts non-trivially on sites $x-n+1,\cdots, x$, then by a generalized version of Kennedy's inversion formula \eqref{eq:Kennedy},
\begin{equation}
\begin{split}
    j_{x}^{(n)}=&\sum_{r=0}^\infty\tr_{x+1,\cdots,x+n-2+r}\frac{d\rho_{x+r}^{(n)}}{dt}\\
    =&\sum_{r=0}^\infty(\tr_{x+1}\mathcal{T})^r\tr_{x+1}\frac{d\rho_{x}^{(n)}}{dt}\\
    =&(1-\tr_{x+1}\mathcal{T})^{-1}\tr_{x+1}\frac{d\rho_{x}^{(n)}}{dt}\equiv\mathcal{P}^{-1}[\frac{d\rho_{x}^{(n)}}{dt}].
\end{split}
\end{equation} We do not need to worry about the convergence of the infinite sum because it terminates due to the cyclic property of trace after it is taken over the entire support of $[H,\rho^{(n)}_x]$. 

One can also introduce superoperators that correspond to a scaling transformation
\begin{equation}
    \mathcal{D}=x(\mathcal{T}^{-1}-1),
\end{equation}and a special conformal transformation~\cite{DiFrancesco:1997nk}
\begin{equation}
    \mathcal{K}=x(x-1)\mathcal{T}^{-1}(\mathcal{T}^{-1}-1).
\end{equation}
Then using the canonical commutation relation of finite difference $[\mathcal{T}-1,x\mathcal{T}^{-1}]=1$, they can be shown to obey the commutation rules of an $\mathfrak{sl}_2$ algebra 
\begin{equation}
        [\mathcal{D},\mathcal{P}]=\mathcal{P},\quad
        [\mathcal{D},\mathcal{K}]=-\mathcal{K},\quad
        [\mathcal{K},\mathcal{P}]=2\mathcal{D}.
\end{equation}Together with $\mathcal{D}[H]=H$, $\mathcal{K}[H]=2B$, and $\mathcal{P}[B]=H$, they belong to a (1+1)D lattice conformal algebra.\footnote{In addition, we also have $\mathcal{D}[B]=\frac{1}{2}\mathcal{D}\big[\mathcal{K}[H]\big]=\frac{1}{2}[\mathcal{D},\mathcal{K}][H]+\frac{1}{2}\mathcal{K}\big[\mathcal{D}[H]\big]=0$.}

\section{The generalized Reshetikhin condition}\label{sec:general}

One outdated counterexample to conserved charges being generated by the boost operator defined above is the Hubbard model. As noted since the early days, combining a charge and a spin sector, it is not a `relativistic' integrable model, meaning the $R$-matrix associated with its Hamiltonian depends on both of the spectral parameters, instead of just the difference of two `rapidities'. Put differently, the gapless modes in the two sectors can propagate at different speeds, which necessarily breaks the lattice Lorentz invariance that the boost operator derived its name from. Yet, a ladder operator that generalizes the boost operator has been found for such models~\cite{PhysRevLett.86.5096}, by expanding the $R$-matrix in terms of the difference between the two spectral parameters, such that the coefficients become functions of one of them
\begin{equation}
	\check{R}_{x,x+1}(\mu,\nu)=1_{x,x+1}
    +\sum_{n=1}^\infty \frac{(\mu-\nu)^n}{n!} \check{R}_{x,x+1}^{(n)}(\nu).
	\label{eq:nonRmat}
\end{equation}
As a result, the Hamiltonian and its constant shift also belong to a class labeled by one of the spectral parameters $\check{R}_{x,x+1}^{(1)}(\nu)=h_{x,x+1}(\nu)+c(\nu)1_{x,x+1}$. The general YBE in braided form becomes
\begin{equation}
\check{R}_{x,x+1}(\nu,\lambda)\check{R}_{x-1,x}(\nu, \mu)\check{R}_{x,x+1}(\lambda,\mu) =\check{R}_{x-1,x}(\lambda,\mu)\check{R}_{x,x+1}(\nu, \mu)\check{R}_{x-1,x}(\nu,\lambda).
\label{eq:nonYBE}    
\end{equation}Taking $\mu=\nu$, it implies the unitarity
\begin{equation}
    \check{R}_{x,x+1}(\mu, \nu)\check{R}_{x,x+1}(\nu, \mu)=1_{x,x+1}.
\end{equation}Like before, unitarity provides a way to determine even order terms in \eqref{eq:nonRmat} from lower order ones. In particular, $\check{R}_{x,x+1}^{(2)}(\nu)=\big(\check{R}_{x,x+1}^{(1)}(\nu)\big)^2+\partial_\nu \check{R}_{x,x+1}^{(1)}(\nu)$. 

The coefficient of $\mu\nu^2$ in \eqref{eq:nonYBE} (after Taylor expanding $\check{R}^{(n)}(\lambda)$ around $\lambda=\mu$) gives the generalized Reshetikhin condition~\cite{PhysRevLett.86.5096}
\begin{equation}
        \left[h_{12}+h_{23},\left[h_{23},h_{12}\right]\right]+\left[h_{12},h'_{23}\right]+\left[h_{23}, h'_{12}\right]+\frac{1}{2}\left([h_{12},h'_{12}]+[h_{23},h'_{23}]\right)=Y_{12}-Y_{23}.
    \label{eq:gRc}    
\end{equation}where $Y_{12}=\check{R}_{12}^{(3)}(\mu)-h^3_{12}(\mu)-\partial_\mu^2 h_{12}(\mu)-\frac{3}{2}\{h_{12}(\mu),\partial_\mu h_{12}(\mu)\}$.\footnote{The apparent difference from Eq.~(12) in Ref.~\cite{PhysRevLett.86.5096} is the result of extra minus signs in the odd order terms of the definition \eqref{eq:nonRmat}.} This can be viewed as an equation of an unknown matrix that is the first order derivative of the Hamiltonian density. 

In order to see how $h'_{12}(0)=\partial_\mu h_{12}(\mu)|_{\mu=0}$ could possibly be solved, we can parametrize again the Hamiltonian in terms of $\mathfrak{su}(n)$ operators in Sec.~\ref{sec:sl}. However, since \eqref{eq:sunham} is an outcome of basis transformation from \eqref{eq:Hamiltonian}, we cannot assume that the bilinear term in $h'_{j,j+1}$ is still diagonal in that basis. Moreover, as $h'_{12}(0)$ does not describe a physical interaction, it does not even have to be symmetric about the degrees of freedom at site $j$ and $j+1$. So, its most general form is
\begin{equation}
    h'_{j,j+1}=p_{\alpha\beta} T^\alpha_j\otimes T_{j+1}^\beta+q_\alpha T_j^\alpha +r_\alpha T_{j+1}^\alpha +c'.
\end{equation}The additional terms in \eqref{eq:gRc} compared to \eqref{eq:Reshetikhin} are given by
\begin{equation}
    \begin{split}
    [h_{12}, h'_{23}]
        =&ia_{\alpha} p_{\delta\gamma}f_\beta^{\alpha\delta}T^\alpha_1 T^\beta_{2} T^\gamma_3-ia_{\alpha } q_\gamma f_\gamma^{\alpha\beta}T^\alpha_1 T^\beta_{2}+i b_\gamma p_{\delta\beta}f_\alpha^{\gamma\delta}T^\alpha_{2} T^\beta_3+ib_\beta q_\gamma f_\alpha^{\beta\gamma} T_2^\alpha, \\
        [h_{23},h'_{12}]
        =&ip_{\alpha\delta}a_{\gamma}f_\beta^{\gamma\delta}T^\alpha_1 T^\beta_{2} T^\gamma_3+ib_{\gamma} p_{\alpha\delta} f_\beta^{\gamma\delta}T^\alpha_1 T^\beta_2+ia_{\beta }r_\gamma f_\gamma^{\alpha\beta}T^\alpha_2 T^\beta_{3}+ib_\beta r_\gamma f_\alpha^{\beta\gamma} T_2^\alpha, \\
        [h_{12}, h'_{12}]
        =&i\left(a_{\gamma} p_{\epsilon\delta}(f_\alpha^{\gamma\epsilon}d_\beta^{\gamma\delta}+f_\beta^{\gamma\delta}d_\alpha^{\gamma\epsilon})+(a_\beta q_\gamma -a_\alpha r_\gamma) f^{\alpha\beta}_\gamma+b_\gamma (p_{\delta\beta}f_\alpha^{\gamma\delta}+ p_{\alpha\delta}f_\beta^{\gamma\delta}) \right) T^\alpha_1 T^\beta_{2}\\
        &+ib_{\beta }  f_\alpha^{\beta\gamma}(q_\gamma T^\alpha_1 +r_\gamma T^\alpha_{2}).
    \end{split}
\end{equation}So the generalized Reshetikhin condition is equivalent to
\begin{equation}
    \begin{aligned}
        ia_{\alpha} p_{\delta\gamma}f_\beta^{\alpha\delta}+ip_{\alpha\delta}a_{\gamma}f_\beta^{\gamma\delta}&=u_{\alpha\beta\gamma}-u_{\gamma\beta\alpha},&\forall \alpha,\beta,\gamma,\\
        ia_{\gamma} p_{\epsilon\delta}(f_\alpha^{\gamma\epsilon}d_\beta^{\gamma\delta}+f_\beta^{\gamma\delta}d_\alpha^{\gamma\epsilon})+2i(a_\beta q_\gamma -a_\alpha r_\gamma)  f^{\alpha\beta}_\gamma&+2ib_\gamma (p_{\delta\beta}f_\alpha^{\gamma\delta}+p_{\alpha\delta}f_\beta^{\gamma\delta}) \\ &=x_{\alpha\beta}+v_{\alpha\beta}-x_{\beta\alpha}-v_{\beta\alpha}
        ,\quad &\forall \alpha,\beta,\\
        b_\beta (q_\gamma +r_\gamma)f_\alpha^{\beta\gamma}&=0,  &\forall \alpha.\\
    \end{aligned} \label{eq:generalizedcondition}
\end{equation}In the first two equations, the symmetry about exchanging the first and last indices of the RHS can only be matched by the LHS if $p_{\beta\alpha}=-p_{\alpha\beta}$ and $q_\alpha=-r_\alpha$, which can be verified for the Hubbard model \cite{PhysRevLett.86.5096}. The antisymmetry make the third equation trivially satisfied, and the first equation for $\alpha\ne \gamma$ and the second equation for $\alpha\ne\beta$ contains $n^2(n^2-1)(n^2-2)/2$ equations for $n^2(n^2-1)/2$ unknown variables, which are $p_{\alpha\beta}$ for $\alpha\ne\beta$, and $q_\alpha$. Since we are considering the generalized Reshetikhin condition, the RHSs of these equations cannot all be zero, otherwise the original Reshetikhin condition would be satisfied. So this is an inhomogeneous linear equation system that either has a unique solution or no solution. If there is no solution, it is a clear sign that integrability in the Yang-Baxter sense is not present. Yet when there is a solution, a generalized version of the higher order conditions \eqref{eq:highercurrent} still needs to be checked to conclusively claim integrability.\footnote{Two comments are in order. First, this serves as a way to determine $h'(0)$ without knowledge of the $R$-matrix. This was pointed out in a comment following Eq. (16) in Ref.~\cite{PhysRevLett.86.5096} as impossible without knowledge of the whole class of Hamiltonians to which the Hubbard model belongs. The more accurate statement in light of Kennedy's method would be that $h'(\mu)$ cannot be determined without knowing the $R$-matrix. Second, the above statement is in no contradiction to Theorem 13 of Ref.~\cite{hokkyo2025integrabilitysingleconservationlaw}, which assumes that the $R$-matrix of an integrable Hamiltonian is known. For non-relativistic integrable Hamiltonians, this happens extremely rarely, and has always been considered a great achievement when it does \cite{PhysRevLett.56.2453,RmatforHubbard}. In that case, a full class of Hamiltonians parametrized by the spectral parameter is already given. However, as an integrability test, the starting point here is a generic Hamiltonian isolated in the parameter space.}

Since the section started with a general form of the YBE, without assuming it to be of difference form, the generalized Reshetikhin condition could even be conjectured as a necessary condition for quantum integrability, at least in the YBE sense. However, to bootstrap the full parameter dependent integrable class of Hamiltonians $H(\mu)$ therefrom seems unlikely, not to mention the $R$-matrix, as all that we could solve in the above procedure is $\partial_\mu h(\mu)|_{\mu=0}$ and $\check{R}_{12}^{(3)}(0)-\partial_\mu^2 h_{12}(\mu)|_{\mu=0}$.

\section{Conclusion}\label{sec:conclusion}

The main result of the article is a bootstrap program for finding the $R$-matrix of a relativistic integrable Hamiltonians. In particular, explicit forms of the higher order Reshetikhin conditions \eqref{eq:highercurrent} has been obtained. Regardless of the order, they all depend on three-local operators, and should be simpler observables to study than the conserved currents of a generalized Gibbs ensemble \cite{PhysRevX.10.011054,PhysRevLett.125.070602}. How Reshetikhin's condition is supposed to imply \eqref{eq:highercurrent} and hence the YBE in practice remains an unanswered question. Future pursuit in this direction would not only further shed light upon how the three-local charge conservation implies existence of all higher order charges, but also ultimately address the question of whether quantum integrability happens always in the YBE sense. 

Instead of being considered as conserved charges that protect the integrability of the Hamiltonian, the charges should be treated on the same footing as collectively a class of integrable Hamiltonians that are mutually conserved charges of one another. For the non-relativistic case, each of Hamiltonians in this countable class labeled by an integer that is the order of the charge, is in turn a continuous class of integrable Hamiltonians labeled by one of the spectral parameters of their $R$-matrix.

One possible direction for extension is if instead of $[H,[H,B]]=0$, one has $\ad_H^nB\ne 0$ and $\ad_H^{n+1}B=0$, for $n\ge2$. In that case, $B(t)$ would still be a time-dependent symmetry potentially good for generating a hierarchy of conserved charges, except its explicit form would contain operators with support up to $n+2$. The first conserved charge would instead become $\ad_H^n B$. In the language of mastersymmetry, $B$ would be an $H$-mastersymmetry of degree $n$ \cite{timedependent}. Perhaps it makes sense to find an inverse of the boost operator that generate lower charges from higher ones, which is also consistent with the viewpoint that all the conserved charges are actually the same charge observed in different reference frames.

It should be noted that despite the lack of a counterexample of non-integrable spin chain that satisfies the Reshetikhin condition but not its higher order cousins, it is known in the mathematics literature that one conserved charge does not imply infinite many for non-linear differential equations~\cite{Onesymmetry}. This contrast might hint that the quantum integrable spin chains we are familiar with today might be just a tiny class of more varieties of quantum notions of integrability, as integrable evolution equations with two spatial dimensions is a well studied subject.

\section*{Acknowledgements}
I thank Makoto Yamashita for fruitful discussions, and Henrik Schou Guttesen for carefully reading the manuscript and providing valuable feedback.

\begin{appendix}
\numberwithin{equation}{section}

\section{Proof of canceling surface fluxes} \label{sec:flux}

For simplicity of presentation, in this appendix the shorthand notation $a_n=\check{R}^{(n)}_{12}, b_n=\check{R}^{(n)}_{23}$ is introduced. Expanding the YBE, the coefficient of the $\xi\zeta^{2m}$ term gives
\begin{equation}
\sum_{k=0}^{2m}(-1)^k\binom{2m}{k}(a_kb_1a_{2m-k}-b_{2m-k}a_1b_k) =\sum_{k=0}^{2m}(-1)^k\binom{2m}{k}(b_{2m-k}b_{k+1}-a_{k+1}a_{2m-k}).
\label{eq:YBE12}
\end{equation}
The RHS is in fact the difference of two identical operators acting on two different pairs of neighboring sites:
\begin{equation*}
	\begin{split}
		&\sum_{k=0}^{2m}(-1)^k\binom{2m}{k}b_{2m-k}b_{k+1}\\
		=&\sum_{k=0}^{2m}(-1)^k\binom{2m}{k}b_{k}b_{2m-k+1}\\
		=&\sum_{k=1}^{2m}(-1)^k\left(\binom{2m+1}{k}-\binom{2m}{k-1}\right)b_{k}b_{2m-k+1}+b_{2m+1}\\
		=&\sum_{k=1}^{2m}(-1)^k\binom{2m+1}{k}b_{k}b_{2m-k+1}-\sum_{k=1}^{2m}(-1)^k\binom{2m}{k-1}b_{k}b_{2m-k+1}+b_{2m+1}\\
		=&\sum_{k=1}^{2m}(-1)^k\binom{2m+1}{k}b_{k}b_{2m-k+1}+\sum_{k=0}^{2m-1}(-1)^k\binom{2m}{k}b_{k+1}b_{2m-k}+b_{2m+1}\\
		=&\sum_{k=0}^{2m}(-1)^k\binom{2m}{k}b_{k+1}b_{2m-k},
	\end{split}
\end{equation*}
where in the last step Eq.~\eqref{eq:uniodd} has been used.
The LHS of Eq.~\eqref{eq:YBE12} can also be massaged into the form of Eq.~\eqref{eq:highercurrent} by expanding the terms in the summand containing the largest subscript $2m$ using Eq.~\eqref{eq:unieven}:
\begin{equation*}
\begin{split}
    &\sum_{k=0}^{2m}(-1)^k\binom{2m}{k}(a_kb_1a_{2m-k}-b_{2m-k}a_1b_k)\\ =&\sum_{k=1}^{2m-1}(-1)^k\binom{2m}{k}(a_kb_1a_{2m-k}-b_{2m-k}a_1b_k)+a_0b_1a_{2m}+a_{2m}b_1a_{0}-b_{2m}a_1b_0-b_{0}a_1b_{2m}\\
    =&\sum_{k=1}^{2m-1}(-1)^k\binom{2m}{k}\left(a_k[b_1,a_{2m-k}]+a_ka_{2m-k}b_1-[b_{2m-k},a_1]b_k-a_1b_{2m-k}b_k\right)\\ 
    &+b_1a_{2m}+a_{2m}b_1-b_{2m}a_1-a_1b_{2m}\\
    =&\sum_{k=1}^{2m-1}(-1)^k\binom{2m}{k}\left(a_k[b_1,a_{2m-k}]-[b_{2m-k},a_1]b_k\right)+b_1a_{2m}-a_{2m}b_1-b_{2m}a_1+a_1b_{2m}\\
    =&\sum_{k=1}^{2m-1}(-1)^k\binom{2m}{k}\left([a_k,[b_1,a_{2m-k}]]+[b_1,a_{2m-k}]a_k-[[b_{2m-k},a_1],b_k]-b_k[b_{2m-k},a_1]\right)\\
    &+b_1a_{2m}-a_{2m}b_1-b_{2m}a_1+a_1b_{2m}\\
    =&\sum_{k=1}^{2m-1}(-1)^k\binom{2m}{k}\left([a_k,[b_1,a_{2m-k}]]-[b_k,[a_1,b_{2m-k}]]\right)-b_1a_{2m}-a_{2m}b_1-b_{2m}a_1-a_1b_{2m}\\
    &-\sum_{k=1}^{2m-1}(-1)^k\binom{2m}{k}(a_{2m-k}b_1a_{k}-b_{k}a_1b_{2m-k})\\
    =&\sum_{k=1}^{2m-1}(-1)^k\binom{2m}{k}\left([a_k,[b_1,a_{2m-k}]]-[b_k,[a_1,b_{2m-k}]]\right)\\ &-\sum_{k=0}^{2m}(-1)^k\binom{2m}{k}(a_kb_1a_{2m-k}-b_{2m-k}a_1b_k).
\end{split}
\end{equation*}From this we get
\begin{equation}
\begin{split}
    &\sum_{k=0}^{2m}(-1)^k\binom{2m}{k}(a_kb_1a_{2m-k}-b_{2m-k}a_1b_k)\\ =&\frac{1}{2}\sum_{k=1}^{2m-1}(-1)^k\binom{2m}{k}\left([a_k,[b_1,a_{2m-k}]]-[b_k,[a_1,b_{2m-k}]]\right).
\end{split}
\end{equation}

\section{Explicit examples of the first steps in the bootstrap program}\label{sec:models}

\subsection{Spin-$\frac{1}{2}$ chains} 

The Heisenberg Hamiltonian is given by taking $T^{\alpha}_j=\sigma^\alpha_j$, $a_\alpha=1$, and $b_\alpha=c=0$ in \eqref{eq:sunham}, where the Pauli matrices with structure constants proportional to the Levi-Civita symbol $f^{\alpha\beta}_\gamma=2\epsilon^{\alpha\beta\gamma}$, and $d^{\alpha\beta}_\gamma=0$. In this case, \eqref{eq:tensorcommutator} is specified to the form
\begin{equation*}
	[\sigma_j^\alpha\otimes \sigma_{j+1}^\beta,\sigma_j^\gamma\otimes \sigma_{j+1}^\delta]=2i\left(\epsilon^{\alpha\gamma\epsilon}\delta^{\beta\delta}\sigma^\epsilon_{j}+\epsilon^{\beta\delta\epsilon}\delta^{\alpha\gamma}\sigma^\epsilon_{j+1}\right).
\end{equation*}
Hence for the Heisenberg Hamiltonian,
\begin{align*}
	[h_{12},h_{23}]=&\sigma_{1}^\alpha\otimes [\sigma_2^\alpha,\sigma_2^\beta]\otimes\sigma_{3}^\beta=-2i\epsilon^{\alpha\beta\gamma}\sigma_{1}^\alpha \sigma_2^\beta\sigma_{3}^\gamma,\\
	[h_{12},[h_{12},h_{23}]]=&-2i\epsilon^{\alpha\beta\gamma}[\sigma_{1}^\delta\otimes\sigma_2^\delta,\sigma_{1}^\alpha\otimes\sigma_2^\beta]\otimes\sigma_{3}^\gamma=4\epsilon^{\alpha\beta\gamma}\left(\epsilon^{\beta\alpha\epsilon}\sigma^\epsilon_{1}+\epsilon^{\alpha\beta\epsilon}\sigma^\epsilon_2\right)\sigma_{3}^\gamma\\
	=&8\left(\sigma_{2}^\alpha\sigma_{3}^\alpha-\sigma_{1}^\alpha\sigma_{3}^\alpha\right),\\
	h_{12}^2=&\sigma_{1}^\alpha\sigma_{1}^\beta \otimes \sigma_2^\alpha\sigma_2^\beta=(\delta^{\alpha\beta}+i\epsilon^{\alpha\beta\gamma}\sigma^\gamma_{1})\otimes (\delta^{\alpha\beta}+i\epsilon^{\alpha\beta\gamma}\sigma^\gamma_2)=3-2\sigma^\alpha_{1}\sigma^\alpha_2\\
	=&3-2h_{12},\\
    h_{12}^3=&3h_{12}-2h_{12}^2=7h_{12}-6.
\end{align*}Kennedy's trace trick then tells us that $X_{12}=8h_{12}$, giving 
\begin{equation}
    R^{(3)}_{12}=X_{12}+(h_{12}+c)^3=3(c^2-2c+5)h_{12}+c^3+9c-6.
\end{equation}In this case, the constant $c$ is not going to matter, so we choose $c=0$ to simplify the following calculations. Eq.~\eqref{eq:unieven} further tells us that $\check{R}_{12}^{(4)}=4h_{12}R_{12}^{(3)}-3h_{12}^4=117-84h_{12}$. In this example, it is easy to see that all higher order conditions are automatically guaranteed by the Reshetikhin condition. Indeed, since all higher order coefficients of $\check{R}_{j,j+1}^{(k)}$ and powers of the local Hamiltonian are proportional to the Hamiltonian density itself up to a constant shift. The LHS of \eqref{eq:highercurrent} are always proportional to the LHS of Reshetikhin condition. At $m=2$, \eqref{eq:highercurrent} becomes
\begin{equation}
    384(h_{23}-h_{12})=621(h_{12}-h_{23})-\check{R}_{12}^{(5)}+\check{R}_{23}^{(5)},
\end{equation}giving $\check{R}_{12}^{(5)}=237h_{12}-540$. We will not continue further, but simply point out that this is leading to the familiar result that the $R$-matrix for the Heisenberg Hamiltonian is a linear superposition of the permutation and the identity operator.

Another curious observation that can be made here is
\begin{equation}
    \ad_{h_{12}}^3h_{23}=-\ad_{h_{23}}^3h_{12}=16[h_{12},h_{23}],
\end{equation}known as the Dolan-Grady (DG) relation \cite{PhysRevD.25.1587} satisfied by the Onsager algebra \cite{PhysRev.65.117}.

Finally, symbolic calculation using \texttt{Mathematica} gives the following relation for the XYZ Hamiltonian
\begin{equation}
	[h_{12}+h_{23},[h_{12},h_{23}]]=2(h_{12}^3-h_{23}^3)-2J^2(h_{12}-h_{23}),
\end{equation}which tells us that for the XYZ Hamiltonian $\check{R}^{(3)}_{12}=2(J_x^2+J_y^2+J_z^2) h_{12}-2h_{12}^3$.

\subsection{Isotropic spin-1 chains}\label{sec:spin-1}

The spin-1 representation of SU(2) is generated by 
\begin{equation*}
	S^1= \frac{1}{\sqrt{2}}\begin{pmatrix}
		0 & 1 & 0\\
		1 & 0 & 1\\
		0 & 1 & 0
	\end{pmatrix},\quad 	
	S^2= \frac{1}{\sqrt{2}}\begin{pmatrix}
		0 & -i & 0\\
		i & 0 & -i\\
		0 & i & 0
	\end{pmatrix},\quad 
	S^3= \begin{pmatrix}
		1 & 0 & 0\\
		0 & 0 & 0\\
		0 & 0 & -1
	\end{pmatrix}.
\end{equation*}Unlike the SU(3) model to be discussed in the next section, none of the generators here has non-vanishing 1,3 entry. But this can be compensated by including powers of the generators in the Hamiltonian, such as 
\begin{equation*}
	(S^+)^2=\frac{1}{2}(S^1+iS^2)^2=\begin{pmatrix}
		0 & 0 & 1\\
		0 & 0 & 0\\
		0 & 0 & 0
	\end{pmatrix}.
\end{equation*} Therefore, it is customary to study the integrability of the class of models 
\begin{equation}
	H_1(\theta)=\sum_j\left(\cos\theta \vec{S}_j\cdot\vec{S}_{j+1}+\sin\theta (\vec{S}_j\cdot\vec{S}_{j+1})^2\right),
\end{equation}as any SU(2) isotropic Hamiltonian of spin-$s$ can be expressed in terms of a polynomial of degree $2s$. 
Solution of the Reshetikhin condition recovers the 6 integrable parameters, $\theta=\pm k\pi/4$, for $k=1,2,3$, as recently confirmed by the exclusion of other possibilities~\cite{park2024proof}. 

In order to see the necessity to include the constant $c$ in \eqref{eq:Rmat}, we focus on the case $\theta=-\pi/4$, known as the Takhtajan-Babujian spin-1 model~\cite{Takhtajan,Babujian}. In terms of the Gell-Mann matrices
\begin{equation}
    \begin{split}
        &T^1= \frac{1}{\sqrt{2}}\begin{pmatrix}
		0 & 1 & 0\\
		1 & 0 & 0\\
		0 & 0 & 0
	\end{pmatrix},\quad 	
	T^2= \frac{1}{\sqrt{2}}\begin{pmatrix}
		0 & -i & 0\\
		i & 0 & 0\\
		0 & 0 & 0
	\end{pmatrix},\quad 
	T^3= \frac{1}{\sqrt{2}}\begin{pmatrix}
		1 & 0 & 0\\
		0 & -1 & 0\\
		0 & 0 & 0
	\end{pmatrix},\\    &T^4= \frac{1}{\sqrt{2}}\begin{pmatrix}
		0 & 0 & 1\\
		0 & 0 & 0\\
		1 & 0 & 0
	\end{pmatrix},\quad 	
	T^5= \frac{1}{\sqrt{2}}\begin{pmatrix}
		0 & 0 & -i\\
		0 & 0 & 0\\
		i & 0 & 0
	\end{pmatrix},\\
	&T^6= \frac{1}{\sqrt{2}}\begin{pmatrix}
		0 & 0 & 0\\
		0 & 0 & 1\\
		0 & 1 & 0
	\end{pmatrix}, \quad   T^7= \frac{1}{\sqrt{2}}\begin{pmatrix}
		0 & 0 & 0\\
		0 & 0 & -i\\
		0 & i & 0
	\end{pmatrix},\quad 	
	T^8= \frac{1}{\sqrt{6}}\begin{pmatrix}
		1 & 0 & 0\\
		0 & 1 & 0\\
		0 & 0 & -2
	\end{pmatrix},
    \end{split}
\end{equation}its Hamiltonian after an overall rescaling can be written as
\begin{equation}
    H_\mathrm{TB}=\sqrt{2}H_1(\pi/4)=\sum_j\left(a_{\alpha\beta} T^\alpha_jT^\beta_{j+1}-\frac{4}{3}\right).
\end{equation}The non-vanishing coefficients are
\begin{equation}
\begin{split}
    a_{11}=a_{16}=a_{61}=a_{66}=a_{22}=a_{27}=a_{72}=a_{77}=-a_{44}=1,\\
    a_{33}=-\frac{1}{4},\quad a_{38}=a_{83}=\frac{3\sqrt{3}}{4},\quad a_{88}=\frac{5}{4}.\label{eq:TBcoeff}
\end{split}
\end{equation}The above normalization of the Gell-Mann matrices have the structure constants
\begin{equation}
\begin{split}
    &f^{12}_3=\sqrt{2}, \quad f^{14}_7=f^{15}_6=f^{24}_6=f^{25}_7=f^{34}_5=-f^{36}_7=\frac{\sqrt{2}}{2}, \quad f^{45}_8=f^{67}_8=\frac{\sqrt{6}}{2},\\
    &d^{11}_8=d^{22}_8=d^{33}_8=-d^{88}_8=\frac{\sqrt{6}}{3}, \quad d^{44}_8=d^{55}_8=d^{66}_8=-d^{77}_8=\frac{\sqrt{6}}{6},\\
    &d^{14}_6=d^{15}_7=-d^{24}_7=d^{25}_6=d^{34}_4=d^{35}_5=-d^{36}_6=-d^{37}_7=\frac{\sqrt{2}}{2},
\end{split}
\end{equation}with all the others not related by symmetry to the above being zero.

In this case, the coefficients in \eqref{eq:LHS} become
\begin{equation}
\begin{split}
    u_{\alpha\beta\gamma}=&a_{\eta\zeta}a_{\iota\epsilon}a_{\delta\gamma}f_\theta^{\delta\epsilon}(f_\alpha^{\eta\iota}d_\beta^{\zeta\theta}+f_\beta^{\zeta\theta}d_\alpha^{\eta\iota}),\\
        w_{\alpha\beta}=&a_{\eta\zeta}a_{\gamma\epsilon}a_{\delta\beta}f_\zeta^{\delta\epsilon}f_\alpha^{\eta\gamma},\\
        x_{\alpha\beta}=&a_{\gamma\zeta}a_{\gamma\epsilon}a_{\delta\beta}f_\eta^{\delta\epsilon}f_\alpha^{\zeta\eta},\\
    v_{\alpha\beta}=&y_\alpha=z_\alpha =0.
\end{split}
\end{equation}
Obviously, the constant $c$ will not matter for the Reshetikhin condition, as it disappears upon taking the commutator. However, it will enter the next order conditions \eqref{eq:highercurrent}. In fact, it can be checked using the \texttt{Mathematica} code provided in the supplementary information that at $m=2$, \eqref{eq:highercurrent} can only be satisfied if $c=5\sqrt{2}/3$ in the original Hamiltonian $H_\mathrm{1}(\pi/4)$ or equivalently $h_{j,j+1}=a_{\alpha\beta} T^\alpha_jT^\beta_{j+1}+2$ in $H_\mathrm{TB}$ with the coefficients \eqref{eq:TBcoeff}. This constant shift in the linear term of the Taylor expansion \eqref{eq:Rmat} corresponds to an overall factor that depends on the spectral parameter in the $R$-matrix for this model.

If the higher Reshetikhin conditions and the bootstrap program were used ignoring the constant shift, based on the belief that it should not affect the integrability of an interaction, then one could falsely arrived at the conclusion that an actual integrable Hamiltonian does not satisfy the YBE, despite fulfilling the Reshetikhin condition. The Takhtajan-Babujian model is the perfect example explaining this red herring.

\subsection{SU($N$) chains}

The third example is the anisotropic SU($N$) chain. The isotropic model with the full SU($N$) invariance has been solved by Sutherland using the nested Bethe ansatz \cite{PhysRevLett.20.98,PhysRevB.12.3795}. In the basis
\begin{equation}
    (E^{\alpha,\beta})_{\gamma,\delta}=\delta_{\alpha\gamma}\delta_{\beta\delta},
\end{equation}the Hamiltonian is given by 
\begin{equation}
    H_{\mathrm{SU}(N)}=\sum_j\sum_{\alpha,\beta=1}^N e_j^{\alpha,\beta}\otimes e_{j+1}^{\beta,\alpha}.
\end{equation}Motivated by the partially integrable model studied in Ref.~\cite{PhysRevB.106.134420}, the SU($N$) symmetry can be broken to an $\mathfrak{S}_N$ permutation symmetry by introducing a diagonal potential
\begin{equation}
    H_{\mathfrak{S}_N}=\sum_j\sum_{\alpha=1}^N \left(\sum_{\beta\ne \alpha}e_j^{\alpha,\beta}\otimes e_{j+1}^{\beta,\alpha}+\Delta_\alpha e_j^{\alpha,\alpha}\otimes e_{j+1}^{\alpha,\alpha}\right).
\end{equation}
Reshetikhin's condition is only satisfied if $\Delta_\alpha=\pm 1$. This confirms the non-integrability of the Hamiltonian for generic $\Delta$ shown in Ref.~\cite{PhysRevB.106.134420} by the violation of the YBE.
\end{appendix}

\bibliography{3local.bib}

\end{document}